\documentclass[aps,prl,twocolumn,superscriptaddress]{revtex4-1}%-1}
\usepackage{amssymb}
\usepackage{mathrsfs}
\usepackage{amsfonts, amsmath, amssymb}
\usepackage{graphicx}
\usepackage{dcolumn}
\usepackage{bm}
\newcommand{\braket}[1]{\langle #1 \rangle}

\newcommand{\stst}{\text{ss}}
\usepackage[normalem]{ulem}
\usepackage{color}
\usepackage[utf8]{inputenc}
\usepackage[colorlinks,citecolor=blue,linkcolor=blue,urlcolor=blue]{hyperref}

\newcommand{\sket}[1]{|| #1 \rangle\rangle}

\begin{document}

\title{Steady-state phase transition in one-dimensional quantum contact process}

\author{Lin Shang}
\affiliation{School of Physics, Dalian University of Technology, 116024 Dalian, China}

\author{Shuai Geng}
\affiliation{School of Physics, Dalian University of Technology, 116024 Dalian, China}

\author{Xingli Li}
\affiliation{Department of Physics, The Chinese University of Hong Kong, Shatin, New Territories, Hong Kong, China}

\author{Jiasen Jin}
\email{jsjin@dlut.edu.cn}
\affiliation{School of Physics, Dalian University of Technology, 116024 Dalian, China}
\date{\today}
\begin{abstract}
We investigate the steady-state phases of the one-dimensional quantum contact process model. We present the Liouvillian gap in the thermodynamic limit and uncover the metastability of the system. Exploiting the mean-field approximations with a novel self-consistent condition based on the effective field, we capture the avoid the interference of the metastable state. We show the feature of saddle-node bifurcation of the order parameter revealing the discontinuous phase transition of the steady state and extract the transition point for infinite-size system. We show the monotonic decreasing of the steady-state magnetic susceptibility by the linked-cluster expansion, which does not support the divergence of the correlation length at the vicinity of the transition point. The present results may be tested in the quantum simulator of Rydberg atoms.
\end{abstract}

\maketitle

%\section{Introduction}
%\label{Introduction}
%{\it Introduction.}
The non-equilibrium behavior of quantum many-body systems has attracted significant attention and has been extensively discussed both experimentally and theoretically \cite{Lee2011,Lee2012,Lee2013prl,Carusotto2013,Jin2013,LeBoite2013,Jin2014,Carmichael2015,Nagy2015,Noh2016,Landa2020,weimer2021,rossini2021,LSS2023,fazio2025}.
Recent breakthroughs in quantum experimental technology and advances in precise quantum control have enabled the realization of quantum many-body systems far from equilibrium in highly controlled laboratory situations ~\cite{Baumann2010,Baumann2011,Fitzpatrick2017,Bernien2017,Zhang2017,Collodo2019,Ding2020,wu2024}. It has spurred the the exploration of the non-equilibrium phenomena such as the many-body localization, quantum thermalization and the nonequilibrium phase transitions \cite{Marcuzzi2015,Gutierrez2017,Helmrich2018,Ma2025}.

A notable model of the phase transitions occurring in nonequilibrium is the contact process which deals with a $d$-dimensional lattice \cite{Harris1974,Grassberger1982,Odor2004}. The site in the lattice can be either empty or occupied by an excitation. The dynamics encompasses two competing processes: (i) self-destruction, consisting of spontaneous decay, and (ii) branching (coagulation), for which an empty (occupied) site can become occupied (empty) only if at least one of the neighboring sites is excited. The contact process plays significant role in modeling the disease spreading, sociological dynamics, and information dissemination~\cite{Mollison1977,Grassberger1983,Linder2008,Castellano2009,Kuhr2011,CPE2017}.

However, when the model is extended to the quantum realm, the new features arise due to the interplay between quantum coherent time evolution and dissipation.
Early investigations into the quantum contact process (QCP) focused on its non-equilibrium phase transition and its distinctions from the classical directed percolation class~\cite{Marcuzzi2016,Buchhold2017,Jo2019,Roscher2018,Carollo2019,Gillman2019,Gillman2020,Jo2021}. A remarkable feature of the QCP is that it can show phase transition in one-dimensional system, however the nature of such a phase transition remains controversial~\cite{Carollo2019,Gillman2019,Marcuzzi2016,Buchhold2017,Jo2019,Roscher2018}. Whether the phase transition is continuous or not still remains a matter of debate, and constitutes the main focus of the present work.

In the following, we exploit the single-site and cluster mean-field (CMF) approximations to investigate the emergence of the active phase by systematically including the correlations in the system. We present the Liouvillian spectrum of the system and uncover the existence of the metastability near the transition point. By adopting a self-consistent condition based on a novel effective field scenario, we prevent the system being trapped in the metastable state and highlight the saddle-node bifurcation of the order parameter. This provides evidence that the system exhibits discontinuous jump from the absorbing to active phases.  Moreover we implement the numerical linked-cluster expansion (LCE) to analysis the length scale of correlations to support the CMF results.

%\section{The Model}
%\label{sec:Model}
{\it The Model.}-- To model the one-dimensional QCP, we consider a quantum spin-$\frac{1}{2}$ chain incorporating the local dissipation at each site. The spins can be either in the spin-up state $|\uparrow\rangle$ or the spin-down state $|\downarrow\rangle$ representing the occupied or the empty sites, respectively. The dissipations always incoherently flip the spins down to the $z$-direction modelling the self-destruction. In QCP, the branching and coagulation work coherently which are modelled by the following Hamiltonian (adopt the units of $\hbar=1$ hereinafter),
\begin{equation}
 \hat{H}=\Omega\sum_{j=1}^{L-1}(\hat{\sigma}_{j}^{x}\hat{n}_{j+1}+\hat{n}_j\hat{\sigma}_{j+1}^{x}),
 \label{eq_Hamiltonian}
\end{equation}
where $\hat{\sigma}^{\alpha}_{j}$ ($\alpha=x,y,z$) are the Pauli operators of the $j$th site, and $\hat{n}_{j}=\hat{\sigma}^{+}_{j}\hat{\sigma}^{-}_{j}$ denotes the number operator with $\hat{\sigma}^{\pm}_{j}=(\hat{\sigma}_{j}^{x}\pm i\hat{\sigma}_{j}^{y})/2$ being the raising and lowering operators. The parameter $\Omega$ denotes the rate of coherent branching/coagulation and $L$ is the number of spins in the chain.

\begin{figure}[!t]
  \includegraphics[width=.85\linewidth]{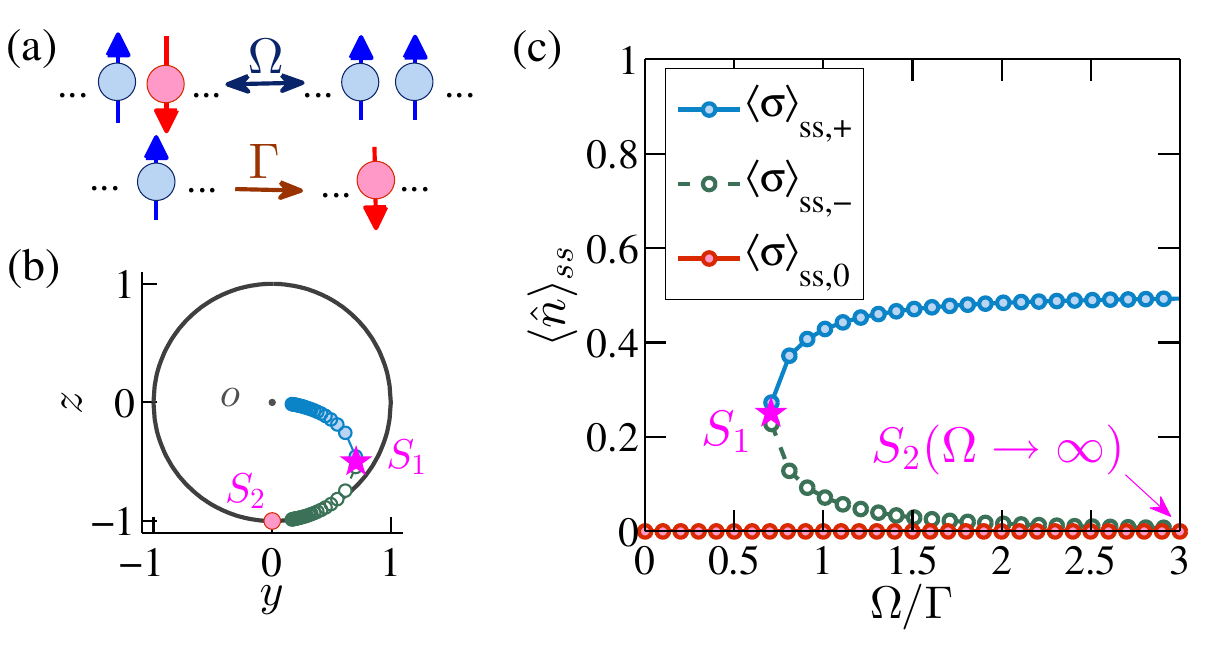}
  \caption{(a) The schematic diagram of the QCP in one dimension. The spin on each site coherently flips with a rate $\Omega$ between the occupied (spin-up) and empty (spin-down) state conditioned on its neighboring sites being occupied. Each spin incoherently decays to the empty state at a rate $\Gamma$. (b) The distribution of the steady states in the $y$-$z$ plane of the Bloch sphere. The blue (green) circles represent the stable (unstable) steady states and the red circles represent the absorbing phase $|\downarrow\rangle$. (c)  The three branches of steady-state population $\braket{n}_{\stst}$ as functions of $\Omega$ within the single-site mean-field approximation.}
  \label{mf_sta}
\end{figure}

Under the Markovian approximation, the time evolution of the density matrix $\rho$ of the system can be described by the following Lindblad master equation with the Liouvillian superoperator ${\cal L}$,
\begin{equation}
\partial_t\rho={\cal L}[\rho]:=-i[\hat{H},\rho]+{\cal D}[\rho].
\label{eq_Lindblad}
\end{equation}
As illustrated in Fig. \ref{mf_sta}(a), the first term on the right-hand side denotes the coherent evolution governed by the Hamiltonian (\ref{eq_Hamiltonian}). The state of a given site flips coherently only when one of its neighboring sites is occupied. While the second term ${\cal D}[\rho]:=\Gamma\sum_{j=1}^L{(\hat{\sigma}_{j}^{-}\rho\hat{\sigma}_{j}^{+}-\frac{1}{2}\{\hat{n}_j,\rho\})}$ denotes the local dissipation on each site with a rate $\Gamma$.

We are interested in the steady state of the system in the thermodynamic limit (TDL), i.e. $\rho_{\text{ss}}=\lim_{t\rightarrow\infty}\lim_{L\rightarrow\infty}{\rho(t,L)}$. It should be emphasized that the two limits do not commute. The averaged steady-state population $\braket{\bar{n}}_{\stst}=\sum_{j=1}^{L}{\braket{\hat{n}_j}_\stst}/L$ serves as the order parameter. There exists a critical value $\Omega_c$, for $\Omega<\Omega_c$ all the sites are in the spin-down state, which is referred to as the absorbing phase with $\braket{\bar{n}}_\stst=0$. Oppositely, for $\Omega>\Omega_c$ the system has the probability to attain a steady state with $\braket{\bar{n}}_\stst\ne0$, referred to as the active phase. As $\Omega$ varies, the steady state of the system may undergo a transition from the absorbing to the active phases. We will investigate the behavior of the order parameter near the phase transition and extract the transition point in TDL.

%\section{The stable steady states under mean-field approximation}
%\label{sec:sss_mf}
{\it Stable mean-field steady states.} We start the discussion with the (single-site) mean-field approximation by making Gutzwiller factorization for the density matrix of the total system $\rho=\bigotimes_{j}\rho_j$.
After substituting the factorized total density matrix into Eq. (\ref{eq_Lindblad}), one yields the following mean-field master equation
\begin{equation}
\partial_t\rho_j=-i[\hat{H}_j^{\text{mf}},\rho_j]+{\cal D}[\rho_j],
\label{eq_ME_mf}
\end{equation}
where $\hat{H}^{\text{mf}}_j=\mathfrak{z}\Omega(\langle\hat{\sigma}^x\rangle\hat{n}+\langle\hat{n}\rangle\hat{\sigma}^x$) is the mean-field Hamiltonian and $\mathfrak{z}=2$ is the coordinate number for one dimensional system. In deriving Eq. (\ref{eq_ME_mf}), the density matrices of each site are assumed to be identical, that is $\langle\hat{O}_j\rangle=\langle\hat{O}\rangle$, $\forall j$, for any local observable on the $j$-th site.

In terms of the Bloch vector, the three sets of physical steady-state solutions are given by
$\braket{\boldsymbol{\hat{\sigma}}}_{\stst,0}=(0,0,-1)$ and $\braket{\boldsymbol{\hat{\sigma}}}_{\stst,\pm}=\left(0,\frac{\Gamma}{2\Omega},-\frac{1}{2}\pm\frac{\Gamma}{2}\sqrt{\frac{1}{\Gamma^2}-\frac{1}{2\Omega^2}}\right)$. The latter solutions, indicating the active state, only exist for $\Omega/\Gamma>1/\sqrt{2}$. Note that the $x$-component of $\braket{\hat{\boldsymbol{\sigma}}}_\stst$ is zero, the steady states locate on the $y$-$z$ plane of the Bloch sphere as shown in the inset of Fig. \ref{mf_sta}(b).

It is worth noting that,  as shown in Fig. \ref{mf_sta}(c), the $\braket{\hat{n}}_\stst$
exhibits a saddle-node bifurcation at the points $S_1$ and $S_2$ with $\Omega/\Gamma$ varies \cite{SM}. The branches of the two active solutions arise at $S_1$ ($\Omega=\Gamma/\sqrt{2}$, $\braket{\hat{n}}_\stst=0.25$), while the unstable active branch is merged to the absorbing branch at $S_2$ ($\Omega/\Gamma\rightarrow\infty$, $\braket{\hat{n}}_\stst=0$). Between $S_1$ and $S_2$ the absorbing and the stable active states coexist. This behavior is reminiscent of the hysteresis in the mean-field analysis of discontinuous phase transition and is significantly different from the supercritical pitchfork bifurcation accompanying the continuous phase transition.

\begin{figure}[!t]
  \includegraphics[width=1\linewidth]{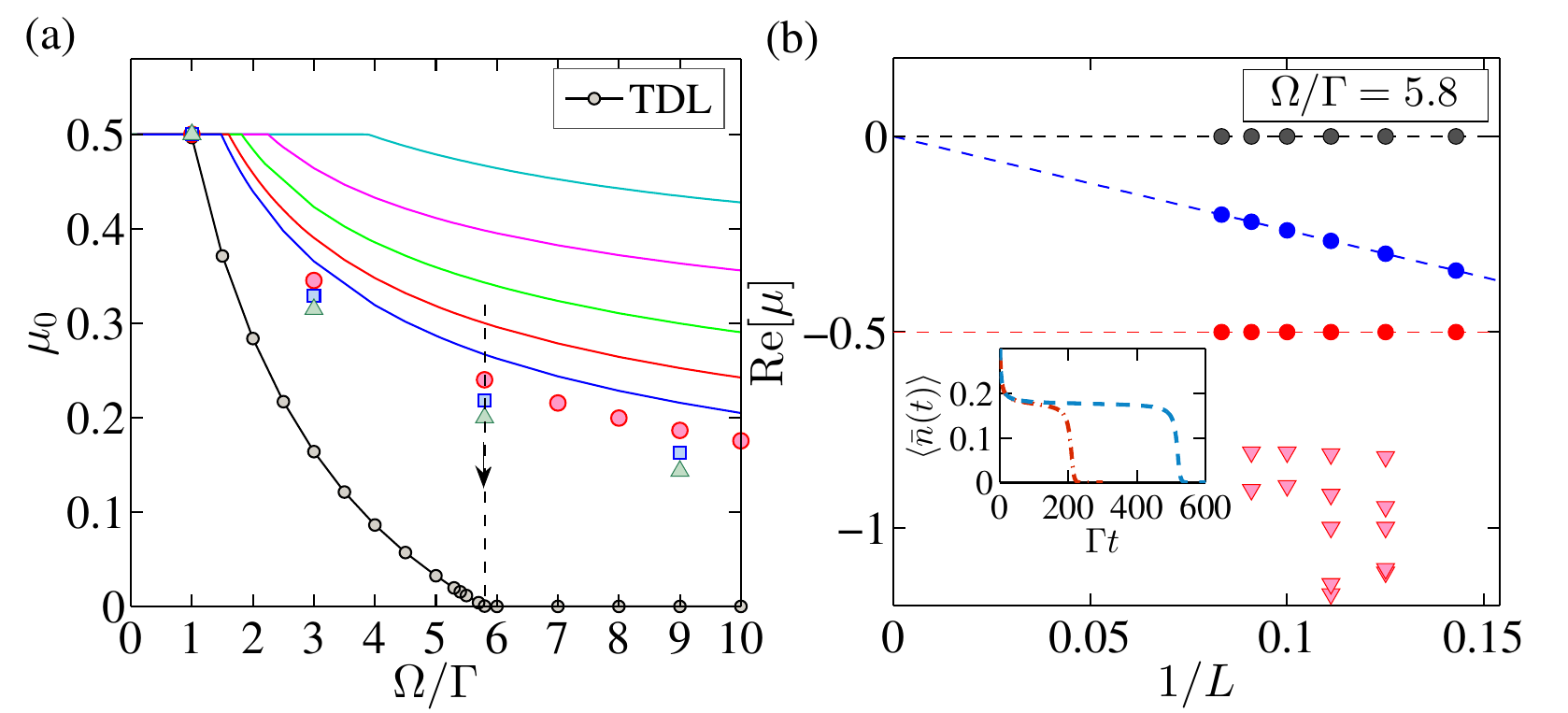}
  \caption{(a) Liouvillian gap. The continuous lines and discrete symbols denote the Liouvillian gap for finite-size system with $L=5,6,$.., and $12$ (from top to bottom). The connected black circles denote the extrapolated Liouvillian gap in the limit of $L\rightarrow\infty$. The gap closes at $\Omega/\Gamma\approx5.83$. (b) The real part of the Liouvillian eigenvalue with different $L$ for $\Omega/\Gamma=5.8$. The black, blue, and red circles represent the zero, first largest negative, and second largest negative real parts of the eigenvalues. The triangles represent the real parts of the rest eigenvalues which are bounded by $\text{Re}(\mu)=-0.5$ (red circles). The inset shows the time-evolution of the averaged population $\braket{\bar{n}(t)}$ with the CMF approximation for $\Omega/\Gamma=5.03$ (red dotted-dashed line) and $5.035$ (blue dashed line). The initial state is the full active state $|\uparrow\uparrow...\uparrow\rangle$ and $L=11$. }
  \label{fig2_Lgap}
\end{figure}

%\section{Liouvillian spectrum and metastability}
%\label{sec:L_spec}
{\it Liouvillian spectrum and metastability.} Although the appearance of the active steady state is observed within the single-site mean-field approximation, actually it survives even if in the full quantum analysis. This is reflected by the Liouvillian spectrum associated to Eq. (\ref{eq_Lindblad}). The real part of the Liouvillian eigenvalue $\mu_i$ is non-positive and characterizes the time-scale of the decay. In particular the eigenstate associated to the zero eigenvalue is the steady state. Define the magnitude of the maximal negative real part as the Liouvillian gap, $\mu_0 = |\max{[\text{Re}(\mu_i)]}|$. The close of the Liouvillian gap indicates the phase transition or bistability \cite{fazio2025,Minganti2018}. In Fig. \ref{fig2_Lgap}(a) the Liouvillian gap as a function of $\Omega$ for various $L$ is shown. The extrapolation of the Liouvillian gap with respect to $L$ indicates the gap closing at $\Omega/\Gamma \gtrsim 5.83$ in TDL, i.e. the coexistence of the active states and the absorbing state \cite{SM}.

Apart from the Liouvillian gap $\mu_0$, the structure of the Liouvillian eigenvalues also reflects the time-scale of relaxation towards the steady state. This is quite relevant to the consuming time in a practical numerical simulation. In Fig. \ref{fig2_Lgap}(b), it is shown the first few largest negative real parts of Liouvillian eigenvalues for $\Omega/\Gamma=5.8$ which is located near the transition point. One can see that the largest negative real part (blue circle) approaches zero as $1/L\rightarrow 0$, as indicated by the arrow Fig. \ref{fig2_Lgap}(a). Moreover the second largest negative real part (red circles) is constant $-0.5$ and it bounds the real parts of all the rest eigenvalues. Namely, the eigenvalues with zero and the largest negative real parts are isolated from the other eigenvalues when the system size becomes larger and larger, signaling a metastability during the time evolution \cite{Macieszczak2016,Rose2016,Macieszczak2021}.

In the inset of Fig. \ref{fig2_Lgap}(b), it is shown that the time evolution of the averaged population $\braket{\bar{n}(t)}$ for $\Omega/\Gamma=5.03$ and $5.035$. Starting with the full active state $|\uparrow\uparrow...\uparrow\rangle$, the $\braket{\bar{n}(t)}$ in both cases decreases fast in the early stage. However it stays in a plateau with finite population for a pretty long time ($> 10^2\Gamma^{-1}$) and then dramatically falls onto the eventual absorbing state. The existence of the metastable state, which is not observed in the (single-site) mean-field approximation, is a consequence of the inclusion of short-range correlations in the system. It reminds us that one should prolong the numerical simulation time to ensure that the system escapes from the metastable state.

\begin{figure}[!t]
  \includegraphics[width=1\linewidth]{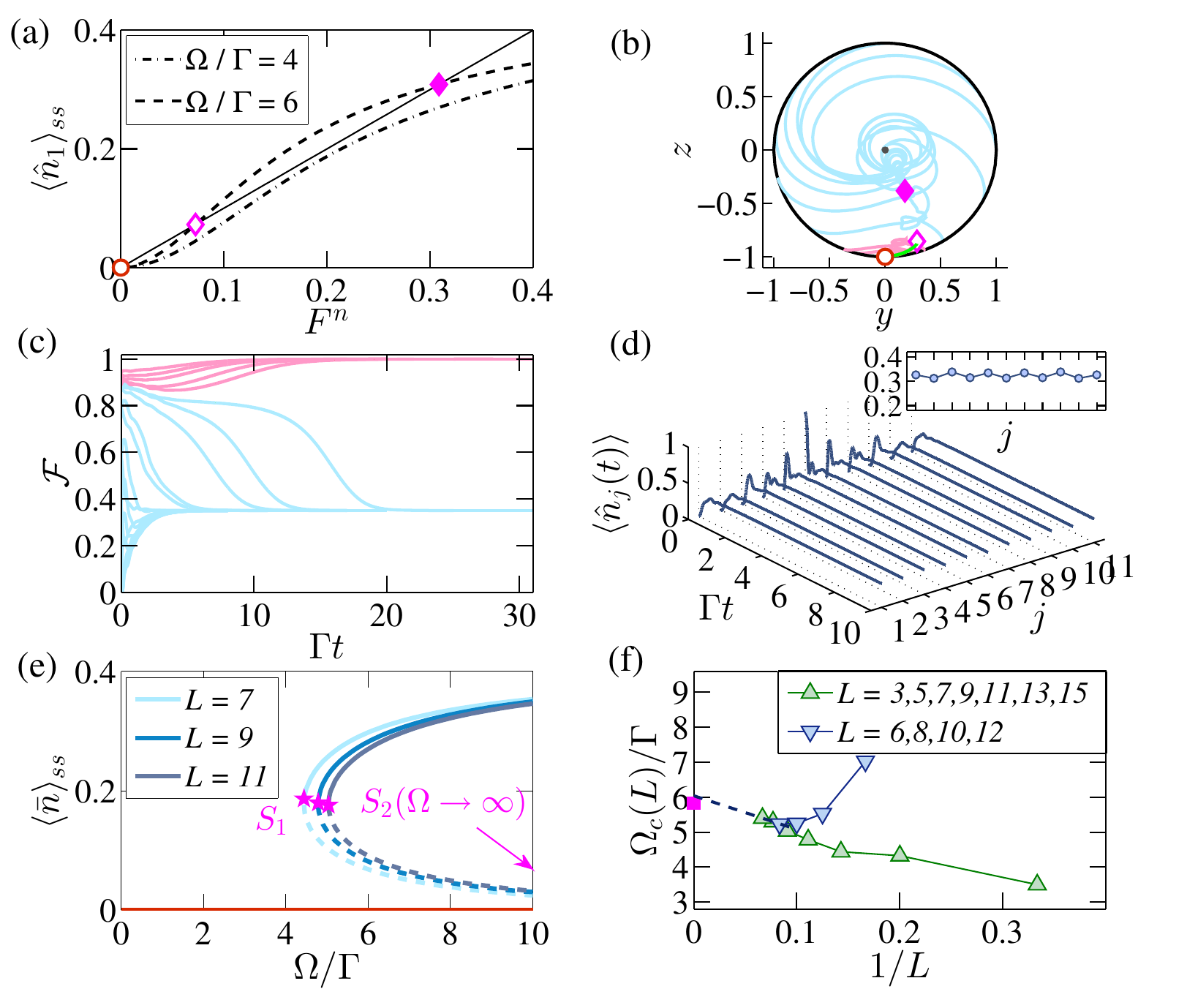}
  \caption{(a) The $\braket{\hat{n}_1}_\stst$ as a function of the effective field $F^n$ for the CMF simulation with $L=7$. The slope of the thin solid line is one. The intersection of the thin solid line and the curve of $\braket{n_1}_{\stst}$ denotes the steady-state solution. The intersection denoted by full (empty) magenta diamond indicates the stable (unstable) steady-state solution. (b) The time-evolution trajectories of the reduced state of the boundary site originating from different initial states $|\psi\rangle=\bigotimes_j{|\psi_j\rangle}$ where the identical $|\psi_j\rangle$s locate at the surface of the Bloch sphere. The diamonds and the circle correspond to the intersections in (a). (c) The time evolution of the fidelity between the system state and absorbing state.  (d) The time evolution of $\langle\hat{n}_j\rangle$ for $\Omega/\Gamma=8$ and $L=11$. The inset of (d) shows the spatial-profile of the steady-state local populations $\langle\hat{n}_j\rangle_{\text{ss}}$. (e) The branches of $\braket{\bar{n}}_\stst$ as a function of $\Omega/\Gamma$ in the CMF approximation with $L=7,9$ and $11$. The upper solid (lower dashed) curves denote the stable (unstable) active states. The red solid line denotes the absorbing state. (f) The resulting CMF transition point $\Omega_c(L) / \Gamma$ for different $L$. The green and blue symbols denote the odd- and even-size clusters, respectively. The dashed line indicates the linear fit $\Omega_c(L)/\Gamma=-9.61/L+6.04$. The square in magenta marks the transition point $\Omega_c/\Gamma = 5.83$ through the Liouvillian gap analysis in Fig. \ref{fig2_Lgap}(a).}
  \label{bcmf_Lv}
\end{figure}

\begin{figure*}[!t]
  \includegraphics[width=0.9\linewidth]{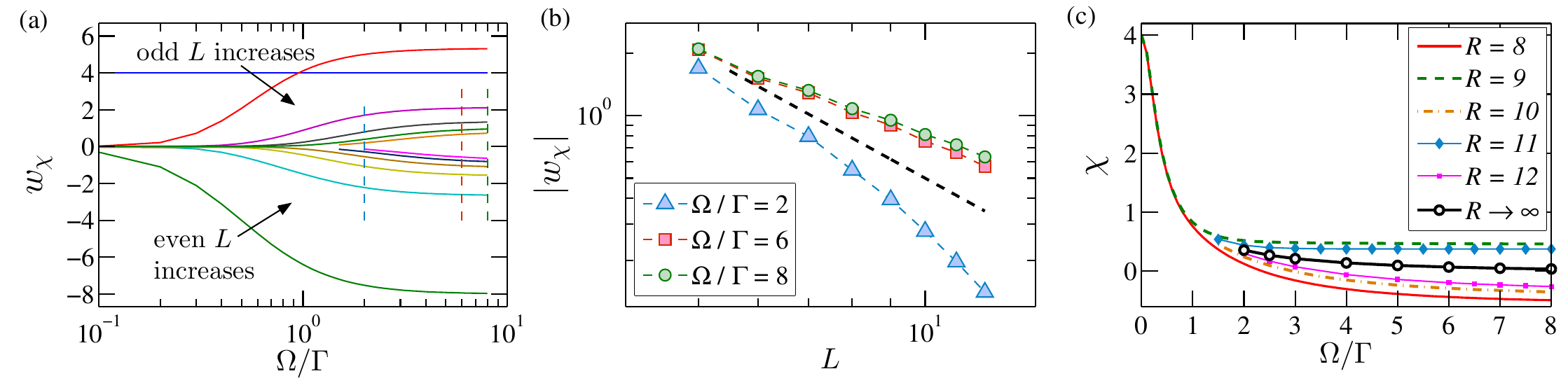}
 \caption{(a) The weight $w_{\chi}(L)$ of the cluster with size $L$ (in logarithmic scale) as a function of $\Omega/\Gamma$. The curves above (blow) the $x$ axis denote the $w_{\chi}(L)$ with odd (even) $L=1,2,3,...,$ and $12$. The two arrows points to the directions along which $L$ increases. (b) The log-log plot of the magnitude $|w_{\chi}(L)|$ scaling with $L$ for various $\Omega/\Gamma$, indicated by the three vertical dashed lines in (a). For $\Omega/\Gamma=2$ the weight scales as $w_{\chi}(L)\approx 8.18e^{-0.34L}$, for $\Omega/\Gamma=6$ the weight scales as $w_{\chi}(L)\approx46.39L^{-1.77}$, and for $\Omega/\Gamma=8$ the weight scales as $w_{\chi}(L)\approx30.13L^{-1.55}$. The dashed line in black represents a power-low function with power $n=-2$ and is plotted as a guide to the eye. (c) The LCE results of $\chi$ with the sum up to various orders of $R$. The connected circles in black represent the extrapolated result with $R\rightarrow\infty$. }
  \label{fig4_LCE}
\end{figure*}

%\section{The effects of short-range correlations}
%\label{sec:CMF}
{\it Effects of the correlations.}
In order to incorporate the effects of correlations in the analysis, we implement the CMF method \cite{Jin2016,Jin2018}. The idea of CMF approximation is to factorize the density matrix of the total system as $\rho=\bigotimes_C{\rho_C}$. Here $\rho_C$ is the density matrix of a cluster composed of $L$ sites. Substituting the factorized density matrix into Eq. (\ref{eq_Lindblad}) and tracing over the clusters other than $C$, one obtains the following CMF master equation,
\begin{equation}
\partial_t\rho_C = -i[\hat{H}_{\text{CMF}},\rho_C]+{\cal D}[\rho_C],
\label{eq_CMF_ME}
\end{equation}
where $\hat{H}_{\text{CMF}}=\hat{H}_C+\hat{H}_{\partial C}(t)$ is the CMF Hamiltonian with $\partial C$ denoting the boundary of $C$. The on-cluster Hamiltonian $\hat{H}_C=\Omega\sum_{j\in C}{(\hat{\sigma}^x_j
\hat{n}_{j+1}+\hat{n}_j\hat{\sigma}^x_{j+1})}$ describes the interaction between the sites inside the cluster $C$ which is treated in a full quantum manner. While the inter-cluster Hamiltonian $\hat{H}_{\partial C}=\Omega\sum_{j\in\partial C, j'\in\partial C'}{(\braket{\hat{\sigma}^x_{j'}}
\hat{n}_{j}+\braket{\hat{n}_{j'}}\hat{\sigma}^x_{j})}$ describes the interaction between cluster $C$ with its neighboring cluster $C'$ via the time-dependent effective fields, where $j, j'=1$ and $L$ being the nearest neighbours but belonging to two adjacent clusters. Due to the translational invariance of the system, the steady-state solution of Eq. (\ref{eq_CMF_ME}) can be determined by the self-consistent conditions $\braket{\hat{O}_{j'}(t\rightarrow\infty)}=\text{tr}[\hat{O}_j\rho_C(t\rightarrow\infty)]$ for $\hat{O}=\hat{n}$ and $\hat{\sigma}^x$.

Here, in order to have a more comprehensive description on the steady-state solutions to Eq. (\ref{eq_CMF_ME}), we reexpress the inter-cluster Hamiltonian as $\hat{H}_{\partial C}=\Omega\sum_{j=1,L}(F^n\hat{n}_j+F^x\hat{\sigma}^x_j)$, where the $F^n$ and $F^x$ play the roles of the effective fields induced by $C'$ and thus parameterized the steady-state density matrix as $\rho_{C,\stst}(F^n,F^x)$. Note that the $x$-component of $\braket{\boldsymbol{\hat{\sigma}}}_{\stst,0}$ and $\braket{\boldsymbol{\hat{\sigma}}}_{\stst,\pm}$ is always zero because the solution with $\braket{\sigma^x}\ne0$ is physically unacceptable \cite{Buchhold2017}. Therefore only $F^n$ is involved in the $\hat{H}_{\partial C}$. The self-consistent condition then requires that $\braket{\hat{n}_j}_\stst=\text{tr}[\hat{n}_j\rho_{C,\stst}(F^n)]$ with $j=1$ or $L$. In this way all the possible self-consistent steady-state solutions can be captured even if it is not stable. Moreover we prevent the system being trapped in the metastable state during the time-evolution.

Fig. \ref{bcmf_Lv}(a) shows the steady-state population of the boundary site $\braket{\hat{n}_1}_\stst$ as a function of $F^n$ for $L=7$. One can see that for $\Omega/\Gamma=4$ the curve of $\braket{\hat{n}}_\stst$ only intersects with the straight line $F^n = \braket{\hat{n}_1}_\stst$ at the origin meaning that the unique steady state is the absorbing state. While for $\Omega/\Gamma=6$ there are two more intersections at $F^n=0.0725$ and $0.3085$ marked by the magenta diamonds. The former (with the derivative greater than one) is an unstable fixed point while the latter indicates the physical steady state. As shown in Fig. \ref{bcmf_Lv}(b), the unstable steady state is brought to the absorbing state by a random perturbation, while stable steady states tends to attract all the time-evolution trajectories originating from different initial states on the surface of the Bloch sphere. The overlap between the initial state and the absorbing state, which is characterized by the fidelity ${\cal F}=\sqrt{\langle\downarrow\downarrow...\downarrow|\rho_C|\downarrow\downarrow...\downarrow\rangle}$, determines which steady state it will evolve. From Fig. \ref{bcmf_Lv}(c), one can find that the initial states with ${\cal F}>0.89$ are attracted by the absorbing state (${\cal F}\rightarrow 1$ in the long-time limit). To verify whether the CMF effectively recovers the translational invariance, we show the time evolution of the local population $\langle\hat{n}_j\rangle$ in the active phase ($\Omega/\Gamma=8$) in Fig. \ref{bcmf_Lv}(d). One observes that starting from the single-excitation state, the spatial profile of $\langle\hat{n}_j\rangle$ becomes homogeneous and the boundary effect is suppressed within the cluster. The stable and unstable steady-state solutions for different $L$ are shown in Fig. \ref{bcmf_Lv}(e). The feature of the saddle-node bifurcation of the steady-state population is preserved in the CMF results. In particular, the saddle-node point $S_2$ remains infinity, while the $S_1$ shifts right towards to the transition point in TDL, which evidences the discontinuous phase transition.

The CMF transition points $\Omega_c(L)$ for different sizes of cluster is shown in Fig. \ref{bcmf_Lv}(f). Due to the strong boundary effect in small clusters, there is an even-odd effect in the behavior of the $\Omega_c(L)$ as a function of $L$. For odd-size clusters, the $\Omega_c(L)$ increases as the $L$ increasing while it exhibits conversely for the even-size clusters. The linear fit of $\Omega_c(L)$ as a function of $1/L$ extrapolates the $\Omega_c(\infty)\approx6.04\Gamma$ which agrees well to the transition point $\Omega_c/\Gamma\approx5.83$ predicted by the Liouvillian spectrum analysis \cite{LinearFit}.

%However for larger $L$ the $\Omega_c(L)$ is expected to converge toward the transition point $\Omega_c/\Gamma\approx5.83$ predicted by the Liouvillian spectrum analysis.

{\it Linked-cluster expansion results.} In order to get further insight about the impact of long-range correlation, we employ the numerical LCE to investigate the magnetic susceptibility $\chi = \lim_{h^x\rightarrow0}{\partial\braket{\bar{\sigma}^y}\stst/\partial h^x}$, which characterizes the linear response of the steady-state magnetization along $y$-direction to a probe field along $x$-direction. The numerical LCE method enables a direct access to the $\chi$ in TDL via a sum of the contributions from the clusters embedded in the system, i.e. $\chi=\sum_{L=1}^{R}{w_{\chi}(L)}$ where $w_{\chi}(L)$ is the weight of the cluster with size $L$ and $R$ is the maximal size in the sum \cite{SM,Biella2018}. The convergence of the sum depends on the typical length scale of the correlations.
In Fig. \ref{fig4_LCE}(a) the weights of clusters with the different sizes are shown, one can see that although the odd- and even-size clusters contribute the positive and negative weights respectively, the magnitude of $|w_\chi|$ decreases as $L$ increasing, revealing that the small clusters contribute dominantly which validates the CMF results. It is interesting that, as $\Omega$ increasing, the weights of larger clusters gradually arise implying the emergence of the active phase. Moreover, as shown in Fig. \ref{fig4_LCE}(b), in the absorbing phase ($\Omega/\Gamma=2$) the $|w_\chi|$ decays exponentially with $L$ increasing. In contrast, in the active phase ($\Omega/\Gamma=8$) a power-law scaling of $|w_\chi|$ with $L$ is observed. The $\chi$ computed by the sum up to $R=12$ is shown in \ref{fig4_LCE}(c). For $\Omega/\Gamma<1.5$ the bare sum already converges at $L\approx9$ which excludes the long-range correlations in the system. For larger $\Omega$, the $\chi$ does not converge even for $R=12$, implying the length scale increases in the active phase. However, the extrapolated $\chi$ shows a monotonic decreasing as $\Omega$ crosses the transition point ( $\Omega\approx 5.83\Gamma$) without the divergence of correlation length at the vicinity of the phase transition \cite{Jin2018}.

%\section{Conclusions}
%\label{sec:conclusions}
{\it Conclusions.} We have investigated the non-equilibrium steady-state phases of the QCP model in one dimension. By means of the mean-field approximation, a saddle-node bifurcation of the order parameter is observed as the controlled parameter $\Omega$ crosses the transition point. We show the closing of Liouvillian gap at $\Omega/\Gamma\approx5.83$ and uncover the existence of the metastability. This suggests one to prolong the numerical simulation time to ensure that the system attains the true steady state. We have proposed a self-consistent condition based on the effective field. It has the advances to avoid the interference of the metastable state and capture all the steady-state solutions. With this novel scenario, the feature of the saddle-node bifurcation remains visible when the effect of correlation is considered, which reveals the discontinuous absorbing-to-active phase transition in one-dimensional QCP. The LCE result shows the monotonicity of magnetic susceptibility without divergence near the transition point. Moreover, the LCE analysis supports that only the short-range correlation contributes dominantly to the steady-state properties and in turn validates the CMF results.

The steady-state phase transition in contact process with both quantum and classical branching and coagulation in the framework of cluster mean-field approximation is the intriguing future direction. In particular it is interesting to extract the critical exponent of the continuous absorbing state phase transition in such hybrid contact process. Along this line, the coherent anomaly analysis basing on the CMF results would play an important role \cite{Suzuki1986,Jin2021}.

\acknowledgments
We warmly thank useful discussions with Prof. Fernando Iemini and Dr. Wen-Bin He. This work was supported by Natural Science Foundation of Liaoning Province No.~2025-MS-009 and the Natural Science Foundation of China under Grants No.~11975064. X. L. is supported by the Hong Kong RGC Early Career Scheme (Grant No. 24308323),  Guangdong Provincial Quantum Science Strategic Initiative GDZX2404004, and the Space Application System of China Manned Space Program.

\widetext
\appendix
\section{Supplemental Material}

\subsection{1. The stability of the single-site mean-field steady-state solutions}

In the single-site mean-field approximation, the Lindblad master equation (2) in the main text is reduced to a master equation regarding to the density matrix of a single site.
The system of the Bloch equations is thus given as follows,
\begin{equation}
\partial_t\braket{\hat{\boldsymbol{\sigma}}} = M\braket{\hat{\boldsymbol{\sigma}}},
\label{eq_Bloch_sys}
\end{equation}
where $\braket{\boldsymbol{\hat{\sigma}}}=(\braket{\hat{\sigma}^x},\braket{\hat{\sigma}^y},\braket{\hat{\sigma}^z})$ is the Bloch vector with $\braket{\hat{\sigma}^\alpha} = \text{tr}(\hat{\sigma}^\alpha\rho_j)$ for $\alpha=x,y$, and $z$. The Jacobian matrix is given as follows,
\begin{equation}
M=\left(
         \begin{array}{ccc}
          -2\Omega\braket{\hat{\sigma}^y}-\frac{\Gamma}{2} & -2\Omega\braket{\hat{\sigma}^x} & 0 \\
         4\Omega\braket{\hat{\sigma}^x} & -\frac{\Gamma}{2} & -2\Omega(2\braket{\hat{\sigma}^z}+1) \\
        0 & 2\Omega(\braket{\hat{\sigma}^z}+1) & 2\Omega\braket{\hat{\sigma}^y}-\Gamma \\
          \end{array}
       \right).
\label{eq_Jacobian}
\end{equation}

By setting $\partial_t\braket{\hat{\boldsymbol{\sigma}}} =0$, one can obtain the steady-state solutions to Eq. (\ref{eq_Bloch_sys}) as
$\braket{\boldsymbol{\hat{\sigma}}}_{\stst,0}=(0,0,-1)$ and $\braket{\boldsymbol{\hat{\sigma}}}_{\stst,\pm}=\left(0,\frac{\Gamma}{2\Omega},-\frac{1}{2}\pm\frac{\Gamma}{2}\sqrt{\frac{1}{\Gamma^2}-\frac{1}{2\Omega^2}}\right)$. The subscript `ss' stands for the steady-state.

The stability of each $\braket{\boldsymbol{\hat{\sigma}}}_{\stst,p}$ (with $p=0, \pm$) is determined by the real parts of the eigenvalues of the corresponding Jacobian matrix (\ref{eq_Jacobian}), denoted as $M_p$. Denote the maximal real part of the eigenvalues of $M_p$ as $\text{Re}(\lambda_p)$. The negative $\text{Re}(\lambda_p)$ means $\braket{\hat{\boldsymbol{\sigma}}}_{\stst,p}$ is stable, otherwise the $\braket{\hat{\boldsymbol{\sigma}}}_{\stst,p}$ is unstable.
\begin{figure}[h]
  \includegraphics[width=0.5\linewidth]{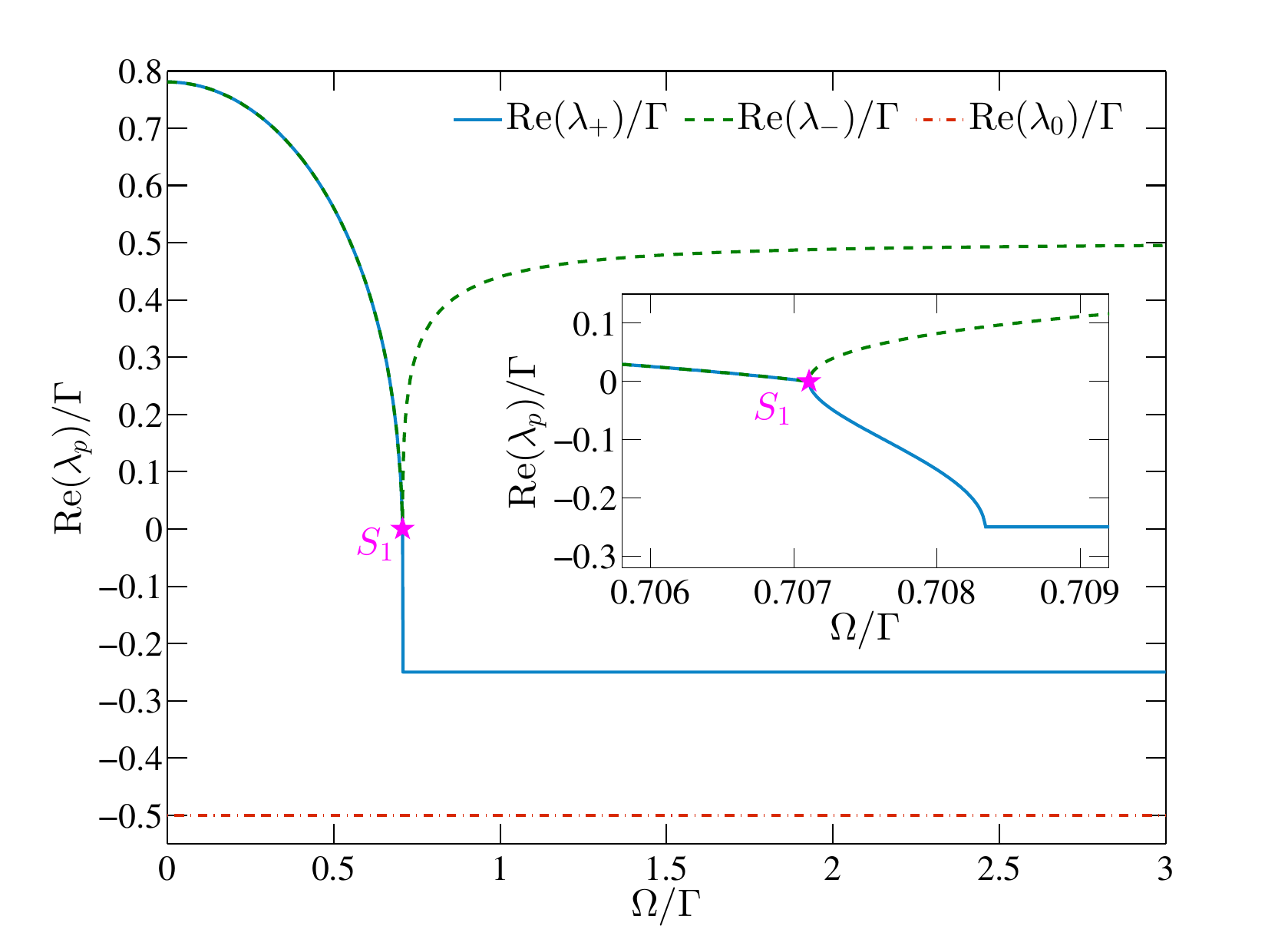}
  \caption{The maximal real part of the eigenvalue of Jacobian Matrix $M_p$ as a function of $\Omega/\Gamma$. The magenta pentagram marks the saddle-node point $S_1$ in Fig. 1 of the main text. Since $\text{Re}(\lambda_0)=-0.5$, the $\braket{\hat{\hat{\sigma}}}_{\stst,0}$ is stable for all $\Omega$. The $\braket{\hat{\hat{\sigma}}}_{\stst,+}$ becomes stable after $\Omega$ crosses the saddle-node point. The inset shows the details of $\text{Re}(\lambda_\pm)$ near the saddle-node point bifurcation.}
  \label{SM_fig1_stamf}
\end{figure}

Substituting the $\braket{\boldsymbol{\hat{\sigma}}}_{\stst}$ into Eq. (\ref{eq_Jacobian}), one obtains the $\text{Re}(\lambda)$ for each  $\braket{\boldsymbol{\hat{\sigma}}}_{\stst}$ as shown in Fig. \ref{SM_fig1_stamf}. One can see that the absorbing steady-state $\braket{\hat{\boldsymbol{\sigma}}}_{\stst,0}$ is always stable indicated by the constant $\text{Re}(\lambda_0)=-\Gamma/2$. The $\braket{\hat{\boldsymbol{\sigma}}}_{\stst,-}$ is always unstable indicated by the positive $\text{Re}(\lambda_-)$. The $\braket{\hat{\boldsymbol{\sigma}}}_{\stst,+}$ is stable for large $\Omega$ and becomes unstable at the saddle-node point $S_1$. Therefore as $\Omega$ increasing the system undergoes a discontinuous transition from the absorbing to the active phases at $\Omega/\Gamma=1/\sqrt{2}$.

\section{2. Liouvillian spectrum analysis}
For the purpose of finding the eigenvalues of Liouvillian superoperator ${\cal L}$ in Eq. (2) of the main text, we perform the vectorization procedure. The density matrix $\rho$ is reshaped into a vector by column-stacking and is denoted as$\sket{\rho}$. In this representation, the identity $\hat{X}\rho\hat{Y}=\hat{Y}^T\otimes\hat{X}\sket{\rho}$ holds for arbitrary operators $\hat{X}$ and $\hat{Y}$, where the superscript `$T$' denotes the matrix transpose \cite{jakob2003}. Then the matrix form of the Liovillian superoperator is given by
\begin{equation}
\mathbb{L} = -i\left( \hat{I}\otimes\hat{H}-\hat{H}^T\otimes\hat{I}\right) + \frac{\Gamma}{2}\sum_{j=1}^L{\left(2\hat{\sigma}^-_j\otimes\hat{\sigma}^-_j-\hat{I}\otimes\hat{\sigma}_j^+\hat{\sigma}_j^--\hat{\sigma}^+_j\hat{\sigma}^-_j\otimes\hat{I}\right)}
\label{eq_L_mat}
\end{equation}
wher $\hat{I}$ is the identity operator. We have adopted $\hat{\sigma}^+_j=(\hat{\sigma}_j^-)^\dagger=(\hat{\sigma}_j^-)^T$ and $\hat{\sigma}^-_j=(\hat{\sigma}_j^-)^*$ in deriving Eq. (\ref{eq_L_mat}).

The eigenvalues $\mu_i$ of the superoperator then can be found by the exact diagonalization of the matrix $\mathbb{L}$ in Eq. (\ref{eq_L_mat}). The steady state of the system is given by the eigenstate associated to the zero eigenvalue of $\mathbb{L}$. Apart from the zero, the real parts of the rest eigenvalues are all negative. It leads to the decay of the amplitudes of rest eigenstates during the time-evolution. In particular, the largest negative real part of the eigenvalues characterize the slowest relaxation scale toward the steady state and is defined as the Liouvillian gap $\mu_0 = |\max{[\text{Re}(\mu_i)]}|$, which is also called the asymptotic decay rate \cite{kessler2012}. Therefore the Liouvilian gap in the thermodynamic limit reveals the features of the steady-state phase transition.

\begin{figure}[h]
  \includegraphics[width=0.9\linewidth]{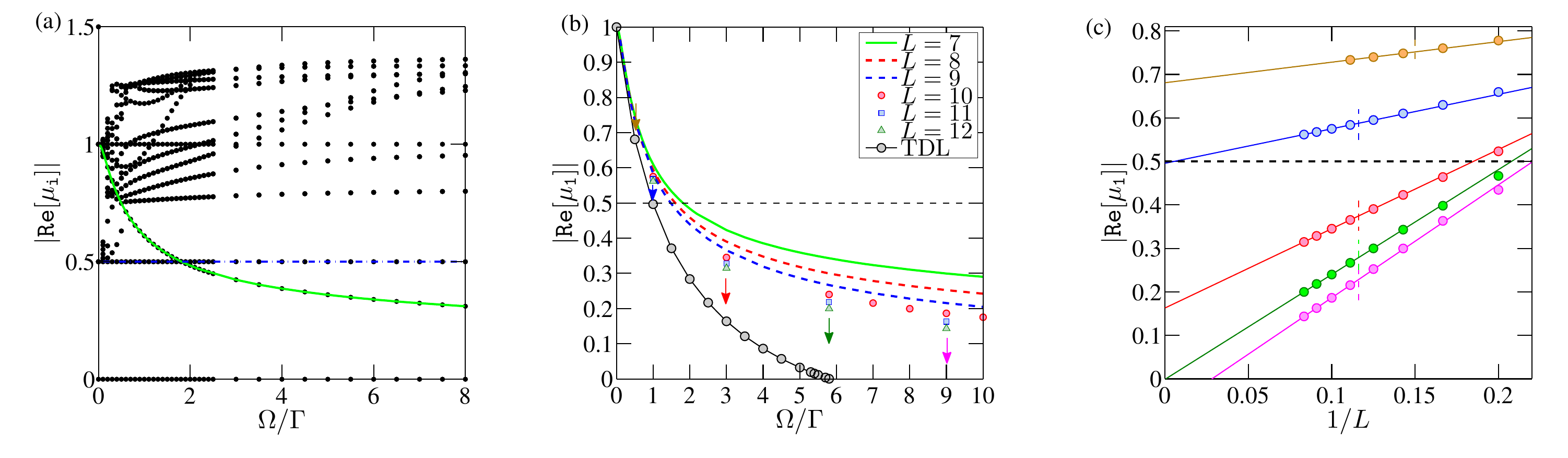}
  \caption{(a) The real part of the low-lying Liouvillian eigenvalue for system size $L=7$ as a function of $\Omega/\Gamma$. The green curve traces the $|\mu_1|$ which is the magnitude of the first non-degenerate non-zero real eigenvalue. The blue horizontal dashed line marks the $|\mu_{\frac{1}{2}}|=0.5$  which is the magnitude of the first two-fold degenerate real eigenvalue. (b) The $\text{Re}[|\mu_1|]$ for different sizes $L$ as a function $\Omega/\Gamma$. The connected circles denote the $\text{Re}[|\mu_1|]$ in TDL. (c) The linear fittings for several typical values of $\Omega/\Gamma$ with respect to $1/L$, From the top to bottom, the lines or symbols represent the data for $\Omega/\Gamma=0.5, 1, 3, 5.8$ and $9$ as indicated by the arrows in (b).}
  \label{SM_fig2_Lgap}
\end{figure}

To this aim we first diagonalize the matrix $\mathbb{L}$ for a finite-size system with $L=7$. The magnitude of the largest forty real parts of the eigenvalues (most of which are degenerate) as a function of $\Omega/\Gamma$ are shown in Fig. \ref{SM_fig2_Lgap}(a). One can see that the gap is basically determined by two eigenvalues, denoted by $\mu_{\frac{1}{2}}$ and $\mu_1$. The eigenvalue $\mu_{\frac{1}{2}}$ is two-fold degenerate and is real with the magnitude $|\text{Re}[\mu_{\frac{1}{2}}]|=0.5$ which is independent of $\Omega/\Gamma$, as marked by the blue horizontal line. While the $\mu_1$ is real and is the first non-degenerate non-zero eigenvalue. It originates from $\mu_1=-1$ at $\Omega=0$ and its magnitude gradually decays to below 0.5 for large $\Omega$, as traced by the green curve.

In Fig. \ref{SM_fig2_Lgap}(b), we show the $|\mu_1|$ as a function of $\Omega/\Gamma$ with $L=7,8,...,12$. For each $\Omega$, we linearly fit the $|\mu_1|$ with respect to $1/L$ and extrapolate the value of $|\mu_1|$ in thermodynamic limit (TDL). In Fig. \ref{SM_fig2_Lgap}(c), the fitting for several typical values of $\Omega$ is displayed. The $|\mu_1(L\rightarrow\infty)|$ are shown by the connected circles in Fig. \ref{SM_fig2_Lgap}(b) (also in Fig. 2(a) of the main text). For small $\Omega$, the real parts of the degenerate eigenvalues between $\mu_{\frac{1}{2}}$ and $\mu_1$ are bounded from below by 0.5. Actually these real parts increases as $\Omega$ increasing. Therefore, the Liouvillian gap in TDL is determined by $\mu_0=\min{(0.5,|\mu_1(L\rightarrow\infty)|)}$.

\section{3. Numerical linked-cluster expansion}
For an extensive observable $\hat{O}=\sum_{j=1}^L{\hat{O}_j}$ of a generic lattice system composed of $L$ sites, the numerical linked-cluster expansion (LCE) enables the direct access to its per-site expectation value in the thermodynamic limit as follows,
\begin{equation}
\lim_{L\rightarrow\infty}{\frac{\braket{\hat{O}}_\stst}{L}}=\sum_{c}{{\mathscr L}(c)\times w_{\hat{O}}(c)},
\label{eq_lce}
\end{equation}
where  the summation is performed over all the linked-cluster $c$ that could appear in the lattice. The multiplicity ${\mathscr L}(c)$ counts the number of ways per site in which cluster $c$ can be embedded on the lattice. The weight $w_{\hat{O}}(c)$ is defined as follows,
\begin{equation}
w_{\hat{O}}(c) = \braket{\hat{O}(c)}_\stst - \sum_{s\subset c}{w_{\hat{O}}(s)},
\label{eq_weight}
\end{equation}
where the summation runs over all the subclusters $s$ in $c$ \cite{tang2013}. The convergence of Eq. (\ref{eq_lce}) is thus guaranteed by the inclusion-exclusion form of Eq. (\ref{eq_weight}).

Regarding the one-dimensional spin-1/2 model in the main text, the clusters $c$ and their subclusters simply reduce to the chains with different sizes $L$ and therefore the multiplicity is ${\mathscr L}(c)=1$ for all clusters $c$. Hence we replace the index $c$ with $L$ in the summation hereafter.

Within the LCE method, the magnetic susceptibility $\chi$ mentioned in the main text can be expressed as follows,
\begin{equation}
\chi=\lim_{h^x\rightarrow0}{\frac{\partial\braket{\bar{\sigma}^y}_\stst}{\partial h^x}}=\lim_{h^x\rightarrow0}{\frac{\partial}{\partial h^x}\sum_{L}{w_{\hat{\sigma}^y}}(L)}=\sum_{L}{\lim_{h^x\rightarrow0}{\frac{\partial w_{\hat{\sigma}^y}(L)}{\partial h^x}}}=\sum_{L=1}^R{w_\chi(L)}.
\label{eq_chi}
\end{equation}
We have defined the weight of $\chi$ as $w_\chi(L)=\lim_{h^x\rightarrow0}{\partial w_{\hat{\sigma}^y}(L)/\partial h^x}$ at the last equality of the above equation. The value of $\chi$ in TDL is obtained with $R\rightarrow\infty$. In a practical computation, one may truncate the sum up to the clusters with finite size $R$. If the sum converges, it means that the contributions from the chains with size larger than $R$ are negligible. This implies that the correlation length of the system is approximately at the order of $R$.

\begin{figure}[h]
  \includegraphics[width=0.95\linewidth]{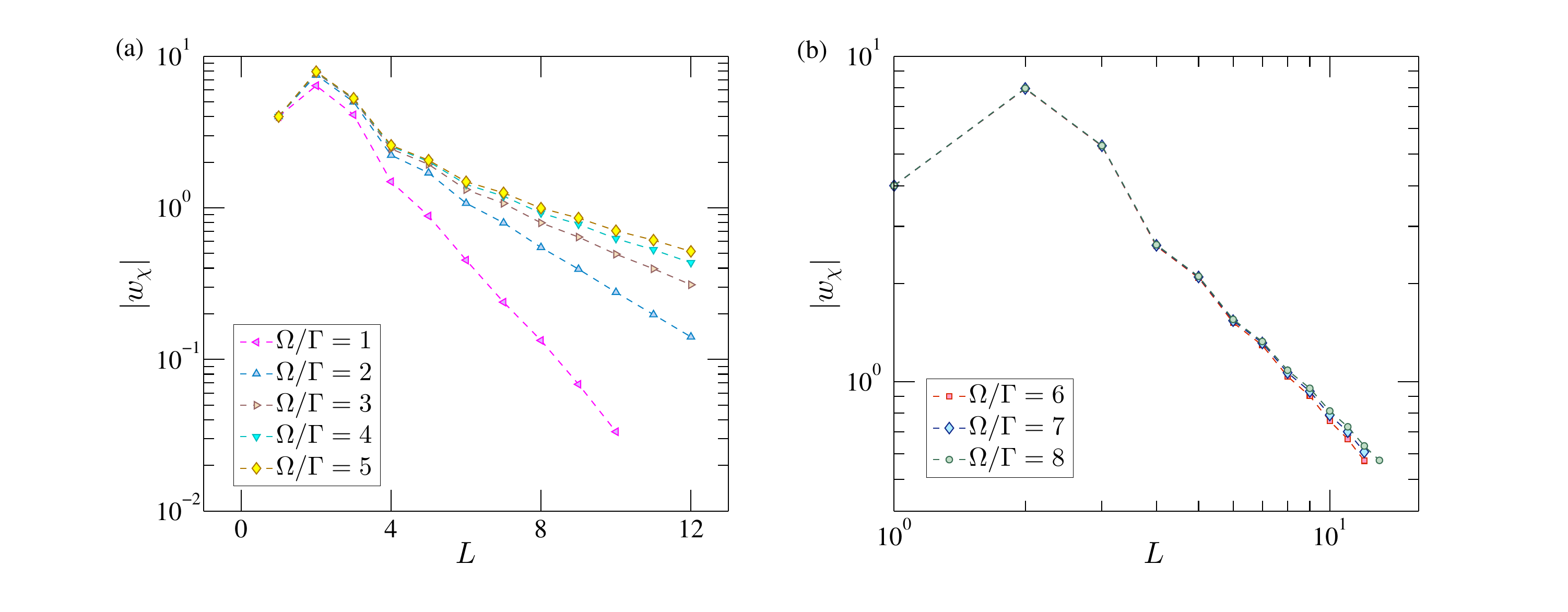}
  \caption{The semi-log plot (a) and log-log plot (b) of $|w_{\chi}(L)|$ as a function of $L$ in the absorbing and active phase, respectively. }
  \label{SM_fig3_weight}
\end{figure}

In the Fig. 4(a) of the main text, it is shown the weight $w_\chi(L)$ as a function of $\Omega/\Gamma$ for various $L$. Here we show a set of more detailed data in Fig. \ref{SM_fig3_weight}. Although, for a given $\Omega$, an obvious even-odd effect of $w_\chi$ on the size $L$ is observed, we can nicely fit the magnitude $|w_\chi(L)|$ for large $L$ with an exponential decay as
\begin{equation}
|w_\chi(L)|=ae^{-bL},
\label{eq_expfit}
\end{equation}
in the absorbing phase and with a power law as
\begin{equation}
|w_\chi(L)|=cL^{-n},
\label{eq_powerfit}
\end{equation}
in the active phase.

As shown in Fig. 4(c) of the main text, the LCE result of $\chi$ has already converged with $R=12$ for $\Omega/\Gamma\lesssim 2$, while for larger $\Omega$ one needs to include the larger $L$ in Eq. (\ref{eq_chi}). In order to obtain the value of $\chi$ in TDL, we fit the $|w_\chi(L)|$ according to Eq. (\ref{eq_expfit}) or (\ref{eq_powerfit}) to extrapolate the value of $w_\chi(L)$ at even larger $L$ ($\ge 13$). Here the $|w_\chi(L)|$ for the largest two $L$ are used for the fitting to minimize the long-tail effect in the power-law scaling.

With the fitted parameters, the value of $\chi$ in the thermodynamic limit is given by
\begin{equation}
\chi=\sum_{L=1}^{R}{w_\chi(L)} + \tilde{\chi}_{\text{abs/act}},
\end{equation}
where in the absorbing phase ($\Omega/\Gamma< 5.83$)
\begin{equation}
\tilde{\chi}_{\text{abs}}=\sum_{L=R+1}^\infty{(-1)^{L-1}ae^{-bL}} =  \frac{a}{1+e^{b}} - \sum_{L=1}^R{ae^{-bL}},
\end{equation}
and in the active phase ($\Omega/\Gamma> 5.83$)
\begin{equation}
\tilde{\chi}_{\text{act}}=\sum_{L=R+1}^\infty{(-1)^{L-1}cL^{-n}} =  c(1-2^{1-n})\zeta(n) - \sum_{L=1}^R{(-1)^{L-1}cL^{-n}},
\end{equation}
with $\zeta(n)$ being the Riemann zeta function.

Basing on the above elaboration, the extrapolated result for $R\rightarrow\infty$ in the Fig. 4(c) of the main text is plotted.

%\begin{thebibliography}{1}
%%% vectorization
%\bibitem{jakob2003}
% M. Jakob and S. Stenholm, Variational functions in driven open quantum systems, Phys. Rev. A {\bf 67}, 032111 (2003).

%%% Liouvilian gap & asymptotic decay rate
%\bibitem{kessler2012}
%E. M. Kessler, G. Giedke, A. Imamoglu, S. F. Yelin, M. D. Lukin, and J. I. Cirac, Dissipative phase transition in a central spin system, Phys. Rev. A {\bf 86}, 012116 (2012).

%%% LCE
%\bibitem{tang2013}
%B. Tang, E. Khatami, and M. Rigol, A short introduction to numerical linked-cluster expansions, Comput. Phys. Commun. {\bf 184}, 557 (2013).
%\end{thebibliography}


\begin{thebibliography}{100}
\bibitem{Lee2011}
T. E. Lee, H. H{\"a}ffner, and M. C. Cross, Antiferromagnetic phase transition in a nonequilibrium lattice of Rydberg atoms, Phys. Rev. A {\bf 84}, 031402(R) (2011).
\bibitem{Lee2012}
T. E. Lee, H. H{\"a}ffner, and M. C. Cross, Collective Quantum Jumps of Rydberg Atoms, Phys. Rev. Lett. {\bf 108}, 023602 (2012).
\bibitem{Lee2013prl}
T. E. Lee, S. Gopalakrishnan, and M. D. Lukin, Unconventional magnetism via optical pumping of interacting spin systems, Phys. Rev. Lett. {\bf 110}, 257204 (2013).
\bibitem{Carusotto2013}
I. Carusotto and C. Ciuti, Quantum fluids of light, Rev. Mod. Phys. {\bf 85} 299 (2013).
\bibitem{Jin2013}
J. Jin, D. Rossini, R. Fazio, M. Leib, and M. J. Hartmann, Photon Solid Phases in Driven Arrays of Nonlinearly Coupled Cavities,  Phys. Rev. Lett. {\bf 110}, 163605 (2013).
\bibitem{LeBoite2013}
A. Le Boit{\'e}, G. Orso, and C. Ciuti, Steady-state phases and tunneling-induced instabilities in the driven dissipative bose-hubbard model, Phys. Rev. Lett. {\bf 110}, 233601 (2013).
\bibitem{Jin2014}
J. Jin, D. Rossini, M. Leib, and M. J. Hartmann, R. Fazio, Steady-state phase diagram of a driven QED-cavity array with cross-Kerr nonlinearities, Phys. Rev. A {\bf 90}, 023827 (2014).
\bibitem{Nagy2015}
D. Nagy and P. Domokos, Nonequilibrium quantum criticality and non-markovian environment: Critical exponent of a quantum phase transition, Phys. Rev. Lett., {\bf115}, 043601 (2015).
\bibitem{Noh2016}
C. Noh and D. G. Angelakis, Quantum simulations and many-body physics with light, Reports on Progress in Physics, {\bf 80}, 016401 (2016).
\bibitem{Landa2020}
H. Landa, M. Schir{\'o}, and G. Misguich, Multistability of Driven-Dissipative Quantum Spins, Phys. Rev. Lett. {\bf 124}, 043601 (2020).
\bibitem{weimer2021}
H. Weimer, A. Kshetrimayum, and R. Or{\'u}s, Simulation methods for open quantum many-body systems, Rev. Mod. Phys. {\bf 93}, 015008 (2021).
\bibitem{rossini2021}
D. Rossini and E. Vicari, Coherent and dissipative dynamics at quantum phase transitions, Physics Reports {\bf 936}, 1 (2021).
\bibitem{LSS2023}
L. da Silva Souza, L. F. dos Prazeres, and F. Iemini, Sufficient Condition for Gapless Spin-Boson Lindbladians, and Its Connection to Dissipative Time Crystals, Phys. Rev. Lett. {\bf 130}, 180401 (2023).
\bibitem{fazio2025}
R. Fazio, J. Keeling, L. Mazza, M. Schir{\'o}, Many-Body Open Quantum Systems, SciPost Phys. Lect. Notes 99 (2025).


\bibitem{Carmichael2015}
H. J. Carmichael, Breakdown of photon blockade: A dissipative quantum phase transition in zero dimensions, Phys. Rev. X {\bf 5}, 031028 (2015).


%% experimental literatures
\bibitem{Baumann2010}
K. Baumann, C. Guerlin, F. Brennecke, and T. Esslinger, Dicke quantum phase transition with a superfluid gas in an optical cavity, Nature {\bf 464}, 1301 (2010).
\bibitem{Baumann2011}
K. Baumann, R. Mottl, F. Brennecke, and T. Esslinger, Exploring symmetry breaking at the dicke quantum phase transition, Phys. Rev. Lett., {\bf 107}, 140402 (2011).
\bibitem{Fitzpatrick2017}
M. Fitzpatrick, N. M. Sundaresan, A. C. Y. Li, J. Koch, and A. A. Houck, Observation of a Dissipative Phase Transition in a One-Dimensional Circuit QED Lattice, Phys. Rev. X {\bf 7}, 011016 (2017).
\bibitem{Bernien2017}
H. Bernien, S. Schwartz, A. Keesling, H. Levine, A. Omran, H. Pichler, S. Choi, A. S. Zibrov, M. Endres, M. Greiner, V. Vuleti{\' c}, and M. D. Lukin, Probing many-body dynamics on a
51-atom quantum simulator, Nature {\bf 551}, 579 (2017).
\bibitem{Zhang2017}
J. Zhang, G. Pagano, P. W. Hess, A. Kyprianidis, P. Becker, H. Kaplan, A. V. Gorshkov, Z.-X. Gong, and C. Monroe, Observation of a many-body dynamical phase transition with a 53-qubit quantum simulator, Nature {\bf 551}, 601 (2017).
\bibitem{Collodo2019}
M. C. Collodo, A. Poto{\v c}nik, S. Gasparinetti, J.-C. Besse, M. Pechal, M. Sameti, M. J. Hartmann, A. Wallraff, and C. Eichler, Observation of the Crossover from Photon Ordering to Delocalization in Tunably Coupled Resonators, Phys. Rev. Lett. {\bf 122}, 183601 (2019).
\bibitem{Ding2020}
D.-S. Ding, H. Busche, B.-S. Shi, G.-C. Guo, and C. S. Adams, Phase Diagram and Self-Organizing Dynamics in a Thermal Ensemble of Strongly Interacting Rydberg Atoms, Phys. Rev. X {\bf 10}, 021023 (2020).
\bibitem{wu2024}
X. Wu, Z. Wang, F. Yang, R. Gao, C. Liang, M. K. Tey, X. Li, T. Pohl, and L. You, Dissipative time crystal in a strongly interacting Rydberg gas, Nat. Phys. {\bf 20}, 1389 (2024).


%%%  applications of noneq. PT
\bibitem{Marcuzzi2015}
M. Marcuzzi, E. Levi, W. Li, J. P. Garrahan, B. Olmos, and I. Lesankovsky, Non-equilibrium universality in the dynamics of dissipative cold atomic gases, New J Phys., {\bf 17}, 072003 (2015).
\bibitem{Gutierrez2017}
R. Guti{\'e}rrez, C. Simonelli, M. Archimi, F. Castellucci, E. Arimondo, D.  Ciampini, M. Marcuzzi, I. Lesanovsky, and O. Morsch. Experimental signatures of an absorbing-state phase transition in an open driven many-body quantum system, Phys. Rev. A {\bf 96}, 041602 (2017).
\bibitem{Helmrich2018}
S. Helmrich, A. Arias, and S. Whitlock, Uncovering the nonequilibrium phase structure of an open quantum spin system, Phys. Rev. A, {\bf 98}, 022109 (2018).
\bibitem{Ma2025}
J.-L. Ma, Z. Guo, Y. Gao, Z. Papi{\'c}, and L. Ying, Liouvillian Spectral Transition in Noisy Quantum Many-Body Scars, Phys. Rev. Lett. {\bf 135}, 180401 (2025).

%%%%  CP  classical
\bibitem{Harris1974}
T. E. Harris, Contact interactions on a lattice, Ann. Probab. {\bf 2}, 969 (1974).
\bibitem{Grassberger1982}
P. Grassberger, On phase transitions in Schl{\"o}gl's second model.  Z. Physik B - Condensed Matter, {\bf 47}, 365 (1982).
\bibitem{Odor2004}
G. {\'O}dor, Universality classes in nonequilibrium lattice systems. Rev. Mod. Phys. {\bf 76}, 663 (2004).
%%% applications of QCP
\bibitem{Mollison1977}
D. Mollison, Spatial contact models for ecological and epidemic spread, J. R. Statist. Soc. B {\bf 39}, 283 (1977).
\bibitem{Grassberger1983}
P. Grassberger, On the critical behavior of the general epidemic process and dynamical percolation, Mathematical Biosciences {\bf 63}, 157 (1983).

%\bibitem{Durrett1994}
%R. Durrett and S. Levin, The importance of being discrete (and spatial). Theoretical Population Biology, 46(3):363{394, 1994.
\bibitem{Linder2008}
F. Linder, J. Tran-Gia, S. R. Dahmen, and H. Hinrichsen, Long-rang epidemic spreading with immunization. J. Phys. A: Math. Theor. {\bf 41}, 185005 (2008).
\bibitem{Castellano2009}
C. Castellano, S. Fortunato, and V. Loreto. Statistical physics of social dynamics, Rev. Mod. Phys. {\bf 81}, 591 (2009).
\bibitem{Kuhr2011}
J.-T. Kuhr, M. Leisner, and E. Frey, Range expansion with mutation and selection: Dynamical phase transition
in a two-species Eden model, New J. Phys. {\bf 13}, 113013 (2011).
\bibitem{CPE2017}
C. P{\'e}rez-Espigares and I. Lesanovsky, Epidemic dynamics in open quantum spin systems, Phys. Rev. Lett. {\bf 119}, 140401(2017).


%%% QCP
\bibitem{Marcuzzi2016}
M. Marcuzzi, M. Buchhold S. Diehl, and I. Lesanovsky, Absorbing state phase transition with competing quantum and classical fluctuations, Phys. Rev. Lett. {\bf 116}, 245701 (2016).
\bibitem{Buchhold2017}
M. Buchhold, B. Everest, M. Marcuzzi, I. Lesanovsky, and S. Diehl, Nonequilibrium effective field theory for absorbing state phase transitions in driven open quantum spin systems. Phys. Rev. B, {\bf 95}, 014308 (2017).
\bibitem{Jo2019}
M. Jo, J. Um, and B. Kahng, Nonequilibrium phase transition in an open quantum spin system with long-range interaction, Phys. Rev. E {\bf 99}, 032131 (2019).
\bibitem{Roscher2018}
D. Roscher, S. Diehl, and M. Buchhold, Phenomenology of first-order dark-state phase transitions, Phys. Rev. A {\bf 98}, 062117 (2018).
\bibitem{Carollo2019}
F. Carllo, E. Gillman, H. Weimer, and I. Lesanovsky, Critical behavior of the quantum contact process in one dimension, Phys. Rev. Lett. {\bf 123}, 100604 (2019).
\bibitem{Gillman2019}
E. Gillman, F. Carollo, and I. Lesanovsky, Numerical simulation of critical dissipative nonequilibrium quantum systems with an absorbing state, New J. Phys. {\bf 21}, 093064 (2019).
\bibitem{Gillman2020}
E. Gillman, F. Carollo, and I. Lesanovsky, Nonequilibrium phase transitions in (1 + 1)-dimensional quantum cellular automata with controllable quantum correlations. Phys. Rev. Lett. {\bf 125}, 100403 (2020).
\bibitem{Jo2021}
M. Jo, J. Lee, K. Choi, and B. Kahng, Absorbing phase transition with a continuously varying exponent in a quantum contact process: A neural network approach, Phys. Rev. Res. {\bf 3}, 013238 (2021).

\bibitem{SM}
See the Supplemental Material for the details of the stability analysis for mean-field results, the Liouvillian spectrum analysis and the elaboration of the numerical linked-cluster expansion, which includes Refs. \cite{jakob2003,kessler2012,tang2013}.

%%% SM-Ref[1]
%%% vectorization
\bibitem{jakob2003}
 M. Jakob and S. Stenholm, Variational functions in driven open quantum systems, Phys. Rev. A {\bf 67}, 032111 (2003).

%%% SM-Ref[2]
%%% Liouvilian gap & asymptotic decay rate
\bibitem{kessler2012}
E. M. Kessler, G. Giedke, A. Imamoglu, S. F. Yelin, M. D. Lukin, and J. I. Cirac, Dissipative phase transition in a central
spin system, Phys. Rev. A {\bf 86}, 012116 (2012).

%%% SM-Ref[3]
%%% LCE
\bibitem{tang2013}
B. Tang, E. Khatami, and M. Rigol, A short introduction to numerical linked-cluster expansions, Comput. Phys. Commun. {\bf 184}, 557 (2013).

\bibitem{Minganti2018}
F. Minganti, A. Biella, N. Bartolo, and C. Ciuti, Spectral theory of Liouvillians for dissipative phase transitions, Phys. Rev. A {\bf 98}, 042118 (2018).

%% metastability
\bibitem{Macieszczak2016}
K. Macieszczak, M. Gu{\c t}{\v a}, I. Lesanovsky, J. P. Garrahan, Towards a Theory of Metastability in Open Quanutm Dynamics, Phys. Rev. Lett. {\bf 116}, 240404 (2016).
\bibitem{Rose2016}
D. C. Rose, K. Macieszczak, I. Lesanovsky, and J. P. Garrahan, Metastability in an open quantum Ising model, Phys. Rev. B {\bf 94}, 052132 (2016).
\bibitem{Macieszczak2021}
K. Macieszczak, D. C. Rose, I. Lesanovsky, and J. P. Garrahan, Theory of classical metastability in open quantum systems, Phys. Rev. Research {\bf 3}, 033047 (2021).

%%%% CMF
\bibitem{Jin2016}
J. Jin, A. Biella, O. Viyuela, L. Mazza, J. Keeling, R. Fazio, and D. Rossini, Cluster Mean-Field Approach to the Steady-State Phase Diagram of Dissipative Spin Systems, Phys. Rev. X  {\bf 6}, 031011 (2016).

\bibitem{Jin2018}
J. Jin, A. Biella, O. Viyuela, C. Ciuti, R. Fazio, and D. Rossini, Phase diagram of the dissipative quantum Ising model on a square lattic, Phys. Rev. B {\bf 98}, 241108(R) (2018).


\bibitem{LinearFit}
The $\Omega_c(L)/\Gamma$ for $L=12,13$ and $15$ are used for the linear fitting to diminish the boundary effect in the finite-size system. We notice that the extrapolated $\Omega_c(\infty)$ is not perfectly accurate due to the residual boundary effects and the fluctuations in calculating the $\Omega_c(L=15)$ with the quantum trajectory method.

%%% LCE
\bibitem{Biella2018}
A. Biella, J. Jin, O. Viyuela, C. Ciuti, R. Fazio, and D. Rossini, Linked cluster expansions for open quantum systems on a lattice, Phys. Rev. B {\bf 97}, 035103 (2018).

%% CAM
\bibitem{Suzuki1986}
M. Suzuki, Statistical Mechanical Theory of Cooperative Phenomena.I. General Theory of Fluctuations, Coherent Anomalies and Scaling Exponents with Simple Applications to Critical Phenomena, J. Phys. Soc. Jpn. {\bf 55}, 4205 (1986).
\bibitem{Jin2021}
J. Jin, W.-B. He, F. Iemini, D. Ferreira, Y.-D. Wang, S. Chesi, and R. Fazio, Determination of the critical exponents in dissipative phase transitions: Coherent anomaly approach, Phys. Rev. B {\bf 104}, 214301 (2021).

\end{thebibliography}
\end{document}